\documentclass[prc,amsmath,twocolumn,showkeys,showpacs,superscriptaddress]{revtex4}
\usepackage{gensymb}
\usepackage{graphicx,color}
\usepackage{amssymb}
\usepackage{enumerate}
\usepackage{verbatim}
\usepackage{natbib}

\begin{document}
  \newcommand {\nc} {\newcommand}
  \nc {\Sec} [1] {Sec.~\ref{#1}}
  \nc {\IR} [1] {\textcolor{red}{#1}} 

\title{Examining the effect of nonlocality in $(d,n)$ transfer reactions}

\author{A.~Ross}
\affiliation{National Superconducting Cyclotron Laboratory, Michigan State University, East Lansing, MI 48824, USA}
\affiliation{Department of Physics and Astronomy, Michigan State University, East Lansing, MI 48824-1321}
\author{L.~J.~Titus}
\affiliation{National Superconducting Cyclotron Laboratory, Michigan State University, East Lansing, MI 48824, USA}
\affiliation{Department of Physics and Astronomy, Michigan State University, East Lansing, MI 48824-1321}
\author{F.~M.~Nunes}
\affiliation{National Superconducting Cyclotron Laboratory, Michigan State University, East Lansing, MI 48824, USA}
\affiliation{Department of Physics and Astronomy, Michigan State University, East Lansing, MI 48824-1321}

\date{\today}

%%%%%%%%%%%%%%%%%%%%%%%%%%%%%%%%%%%%%%%%%%%%%%%%%%%%%%%%%%%%%%%%%%%%%%%%%%%%%%%%%%%%%%%%%%%%%%%%%%%%%%%%%%%%%%%%%%%%%%%%%%%%%%%%%%%

\begin{abstract}
\begin{description}

\item[Background:]  In the last year we have been exploring the effect of the explicit inclusion of nonlocality in (d,p) reactions.
\item[Purpose:] The goal of this work is to extend previous studies to (d,n) reactions, which, although similar to (d,p), have specific properties that merit inspection.  
\item[Method:] We apply our methods (both the distorted wave Born approximation and the adiabatic wave approximation) to 
$(d,n)$  reactions on $^{16}$O, $^{40}$Ca, $^{48}$Ca, $^{126}$Sn, $^{132}$Sn, and $^{208}$Pb at $20$ and $50$ MeV. 
\item[Results:] We look separately at the modifications introduced by nonlocality in the final bound and scattering states, as well as the consequences reflected on the differential angular distributions. The cross sections obtained when using nonlocality explicitly are significantly different than those using the local approximation, just as in (d,p). Due to the particular role of Coulomb in the bound state, often we found the effects of nonlocality to be larger in (d,n) than in (d,p).
\item[Conclusions:] Our results confirm the importance of including nonlocality explicitly in deuteron induced reactions.
\end{description}
\end{abstract}

\pacs{21.10.Jx, 24.10.Ht, 25.40.Cm, 25.45.Hi}

\keywords{elastic scattering, d,n reactions, bound states, transfer reactions, nonlocal optical potentials, Perey effect}

\maketitle

\section{Introduction}

Transfer reactions are of great interest to nuclear structure and nuclear astrophysics, as a means to probe the properties of nuclei and their reactions.
Deuteron induced transfer reactions play a prominent role in our field and are particularly attractive from the theoretical perspective due to the controlled number of channels in the reactions mechanism. Even though deuteron induced single nucleon transfer has been used in our field for several decades, there are still aspects concerning the reaction theory that deserve attention.

Recently we have studied the effects of nonlocality in the optical potentials on (d,p) reactions \cite{Titus_prc2014,Ross_prc2015,Titus_prc2015}. 
In \cite{Titus_prc2014,Ross_prc2015} we focused on nonlocality in the proton channel, namely in the neutron bound state and the proton scattering state. We found that the effects were significant, reducing the magnitude of the wave function in the nuclear interior. For the bound state, due to the normalization condition, the reduction of the magnitude in the nuclear interior resulted in an increase of the asymptotic tail. As a consequence, in general, nonlocality in the bound state and the scattering state produced effects of opposite sign, an increase in the cross section due to the bound state and a decrease due to the scattering state. We found the overall effect to be mostly dominated by the bound state but strongly dependent on the beam energy. In all cases nonlocality was found to be significant.

The original study \cite{Titus_prc2014} used the Perey and Buck nonlocal potential \cite{Perey_np1962}. The  study was repeated for $^{40}$Ca(d,p) reactions \cite{Ross_prc2015} using the dispersive optical potential \cite{Mahzoon_prl2014}. The comparison between results obtained with the Perey and Buck interaction and the DOM interaction demonstrated that, although the magnitude of the effects can be optical potential dependent, they are always significant and need to be carefully considered in the analysis of deuteron induced reactions.

More recently \cite{Titus_prc2015} we have generalized the adiabatic wave approximation \cite{Johnson_npa1974} to include nonlocal interactions. This enabled the inclusion of nonlocality in the deuteron channel, when studying (d,p) reactions. Our systematic study of (d,p) reactions \cite{Titus_prc2015} shows that the effects of nonlocality in the entrance channel are weaker for lighter systems, but can become very important for reactions on heavy targets.
In \cite{Titus_prc2015} we have also investigated the effective method of including nonlocality through an energy shift \cite{Timofeyuk_prl2013}.

Proton transfer (d,n) reactions are an important complementary tool to the neutron transfer (d,p) in studying nuclear structure. These provide important insight into the proton orbitals in the nucleus. These reactions are also pursued for astrophysical reasons. At the energies of astrophysical interest, proton capture rates are unfeasibly low and (d,n) reactions provide a good indirect tool to access the same information (e.g. \cite{Anu_epja2016}).

The (d,n) reactions are experimentally more demanding than (d,p) reactions given the challenges with the detection of the neutron.  Stable beam facilities have carried out a few (d,n) studies throughout the last few decades (e.g. \cite{48Ca_data, 208Pb_data}). In addition, nowadays there is a renewed interest in (d,n) due to the exciting opportunities brought with rare isotope experiments in inverse kinematics. This is demonstrated by the new detector developments that are ongoing in various labs (e.g. LENDA \cite{lenda}, VANDLE \cite{vandle}) but also by the number of (d,n) experiments approved in various PACs (National Superconducting Cyclotron Laboratory alone has 3 such experiments in the books).  
The angular distributions of the outgoing neutron provide angular momentum information on the state populated and are an essential element in the standard analysis. In some recent cases, the challenge of neutron detection has been circumvented by measuring the $\gamma$-rays \cite{Anu_epja2016}, with excellent energy resolution. Then no angular distributions can be extracted, only total cross sections.

On the theoretical side, the same methods that are applicable to (d,p) can usually be immediately applied to (d,n). However the sensitivities to the model space and input parameters are generally not the same. The Faddeev method \cite{Deltuva_prc2009} has recently been applied to (d,n) \cite{Deltuva_prc2015}. As pointed out in that work, (d,n) represented different challenges to (d,p), particularly concerning the handling of the Coulomb through screening. In \cite{Deltuva_prc2015} only  light targets were considered and the aim was mostly to establish whether the Faddeev method could indeed reproduce the existing (d,n) data. 

This work follows naturally from our previous studies \cite{Titus_prc2014,Ross_prc2015,Titus_prc2015}.  Here we concentrate on the effects of nonlocality in (d,n) reactions, using both the distorted wave Born approximation and the adiabatic wave approximation. We summarize briefly the ingredients necessary for the calculation of the cross section in Section II. Next, in Section III,  we present the results for (d,n) on $^{16}$O, $^{40}$Ca, $^{48}$Ca, $^{126}$Sn, $^{132}$Sn and $^{208}$Pb at $E_d=20$ and $50$ MeV as well as a discussion of these results, and a comparison with our previous (d,p) results. Finally, in Section IV we draw our conclusions.

%%%%%%%%%%%%%%%%%%%%%%%%%%%%%%%%%%%%%%%%%%%%%%%%%%%%%%%%%%%%%%%%%%%%%%%%%%%%%%%%%%%%%%%%%%%%%%%%%

\section{Theory and inputs}
\label{theory}

The theory used in our current studies is covered in detail in \cite{Titus_prc2014,Titus_prc2015}. Here we just highlight the various ingredients necessary and the relevant inputs used.

The (d,n) transfer cross section is obtained through the calculation of the T-matrix \cite{book}. The exact T-matrix for the $A(d,n)B$ reaction can be written in the post form as:

\begin{eqnarray}\label{eq-tmatrix}
T=\langle \phi_{pA}\chi^{(-)}_{nB}|V_{np}+\Delta|\Psi^{(+)}\rangle,
\label{eq-tmatrix}
\end{eqnarray}

\noindent where $\phi_{pA}$ is the proton-target bound state, and $\chi^{(-)}_{nB}$ is the neutron scattering state in the exit channel distorted by $U^*_{nB}$. 
The initial state $\Psi^{(+)}$ is the full three-body solution describing the $d+A$ relative motion.
The transition operator is dominated by the NN interaction and the additional  operator $\Delta=U_{nA}-U_{nB}$ has a negligible contribution to the cross section for all cases considered here, with the exception of the lightest target. In this work, we study the effects of the inclusion of nonlocality in the  potentials $U_{nA}$, $U_{pA}$ and $U_{nB}$.

We will first start by focusing on the effects of nonlocality in the final state, which can be done in the distorted wave Born approximation (DWBA).
In the standard formulation of DWBA, the exact three-body initial deuteron state is replaced by the elastic channel:
\begin{eqnarray}\label{tmat-dwba}
T=\langle \phi_{pA}\chi^{(-)}_{nB}|V_{np}| \phi_{d} \chi_{el}    \rangle.
\label{eq-tmatrix}
\end{eqnarray}
Here $\phi_d$ is the deuteron ground state wave function and $\chi_{el}$ corresponds to the solution of an optical model equation with a deuteron optical potential usually fitted to deuteron elastic scattering.

Secondly we will include nonlocality in the entrance channel, for this purpose we will use the finite range ADWA of \cite{Titus_prc2015}. In that method, the T-matrix for the transfer process looks like:
\begin{eqnarray}\label{tmat-adwa}
T=\langle \phi_{pA}\chi^{(-)}_{nB}|V_{np}| \phi_{0} \chi_{ad}    \rangle,
\label{eq-tmatrix}
\end{eqnarray}
where $\phi_{0}$ is proportional to the deuteron bound state wave function and  $\chi_{ad}$ is the scattering solution of a single channel nonlocal equation, where the distorting potential is an adiabatic potential derived from a three-body model for the reaction, and includes deuteron breakup to all orders (see \cite{Titus_prc2015} for more details).

The inputs to our calculations are completely consistent with the calculations in \cite{Titus_prc2014} for the DWBA calculations and with \cite{Titus_prc2015} for the ADWA calculations. This is important as part of the aim of this study is the comparison between (d,p) and (d,n) in what concerns nonlocality.
As in \cite{Titus_prc2014,Titus_prc2015}, the local interactions  for investigating the role of nonlocality are obtained by imposing specific physical constraints. As such, the local interactions reproduce the same elastic scattering as the Perey and Buck interaction, and the local interaction used for the bound states have the same geometry as the real part of the Perey and Buck interaction and reproduce the same binding energy.

%%%%%%%%%%%%%%%%%%%%%%%%%%%%%%%%%%%%%%%%%%%%%%%%%%%%%%%%%%%%%%%%%%%%%%%%%%%%%%%%%%%%%%%%%%%%%%%%%
\section{Results}
\label{results}

\subsection{Nonlocality in the final bound and scattering states}

\begin{table}[b]
\centering
\begin{tabular}{cccccc}
\hline
 			& orbital 		&S$_{p/n}$ (MeV)	& $C_{loc}$ & $C_{nloc}$ & $\Delta(C^2)$\\
\hline
p+$^{16}$O	&d$_{5/2}$	&0.60	&1.476&	1.639&	23.3\\
n+$^{16}$O	&d$_{5/2}$	&4.14	&1.094&	1.286&	38.2\\ \hline
p+$^{40}$Ca	&f$_{7/2}$		&1.09	&32.375&	39.38&	48.0\\
n+$^{40}$Ca	&f$_{7/2}$		&8.36	&3.09&	3.685&	42.2\\ \hline
p+$^{48}$Ca	&f$_{7/2}$		&9.63	&45.029	&56.56	&57.7\\
n+$^{48}$Ca	&p$_{3/2}$		&5.15	&6.367	&7.177	&27.1\\ \hline
p+$^{126}$Sn	&g$_{7/2}$	&7.97	&4446	&6683	&126 \\
n+$^{126}$Sn	&h$_{11/2}$	&5.55	&1.301	&1.468	&27.3\\ \hline
p+$^{132}$Sn	&g$_{7/2}$	&9.67	&4791&	7243		&129. \\
n+$^{132}$Sn	&f$_{7/2}$	&2.37	&1.503&	1.716	&30.4\\ \hline
p+$^{208}$Pb	&h$_{9/2}$	&3.80	&2.32E7	&3.60E7&	139. \\
n+$^{208}$Pb	&g$_{9/2}$	&3.94	&3.46	&3.99&	33.0\\
\hline
\end{tabular}
\caption{Properties of the final bound states: orbital $(\ell,j)$, the separation energy $S_{p/n}$, the ANC for the bound states produced with the local interaction $C_{loc}$ and the nonlocal interaction $C_{nloc}$, and the percent difference between the square of the nonlocal and local ANCs, relative to the nonlocal $\Delta(C^2)$.}
\label{bound}
\end{table}

When considering nonlocality in (d,n) as compared to (d,p) it is important to realize that the final state has in general different quantum numbers, and the transferred nucleon has a different separation energy. In Table \ref{bound} we present the detail of the final proton state for the various reactions to be studied and compare its properties to those neutron states populated in the corresponding (d,p) reaction \cite{Titus_prc2014}. 
We note that in all our calculations we assume a pure single particle structure for the proton states, with spectroscopic factor $S=1$.
Because the reactions we are considering are mostly peripheral, we also look at the single particle asymptotic normalization coefficient (ANC) of the tail of the wave function, as defined in \cite{book}.

One of the important differences between (d,n) and (d,p) reactions is the role of the Coulomb force in the bound state. This aspect, combined with differences in the angular momentum, can change considerably the effect that nonlocality has on the tail of the wave function due to the normalization condition. In Table \ref{bound} we show, for all the single-particle states considered in this study, the ANC obtained when the local interaction is used, the ANC when the nonlocal interaction is used, and the percentage difference between the square of the ANC for the nonlocal interaction and the square of the ANC obtained when the local interaction is used, relative to the nonlocal. For the heavier systems, the percentage difference in the squares of the ANCs is larger for the proton single-particle states than in the neutron single-particle states. The strong Coulomb concentrates the probability at shorter distances, and therefore enhances the effect of nonlocality.

We found in all (d,n) cases studied that the effect of nonlocality in the scattering state was much weaker than that on the bound states. The differences on the scattering state are in all similar to those described in \cite{Titus_prc2014}, namely a reduction of the amplitude in the interior region. As we will demonstrate in the following section, it is the bound state properties that dominate the effect of nonlocality in the (d,n) cross sections.

\subsection{Transfer cross sections}

\begin{figure}[t]
\center
\includegraphics[scale=0.27]{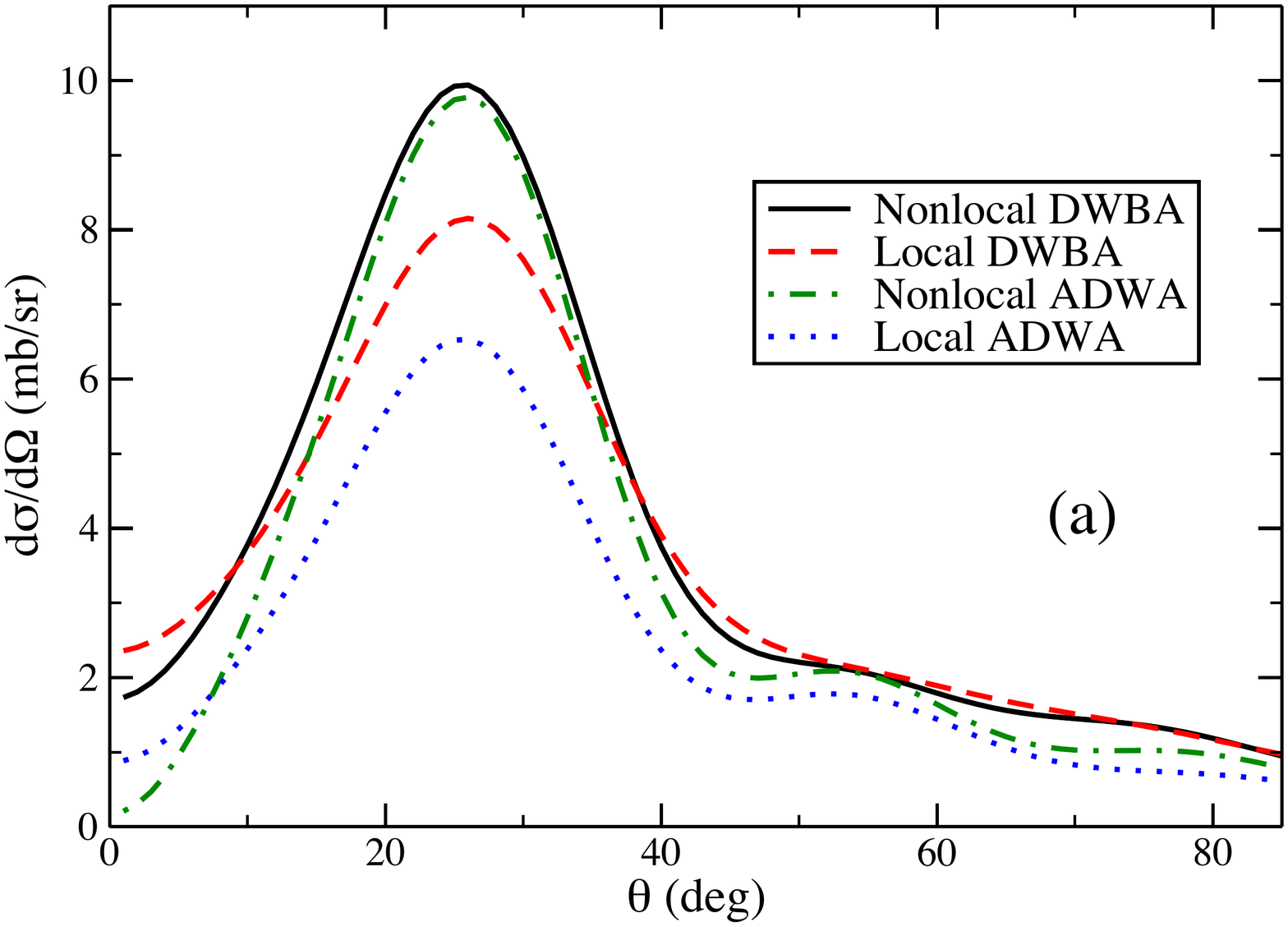}
\includegraphics[scale=0.27]{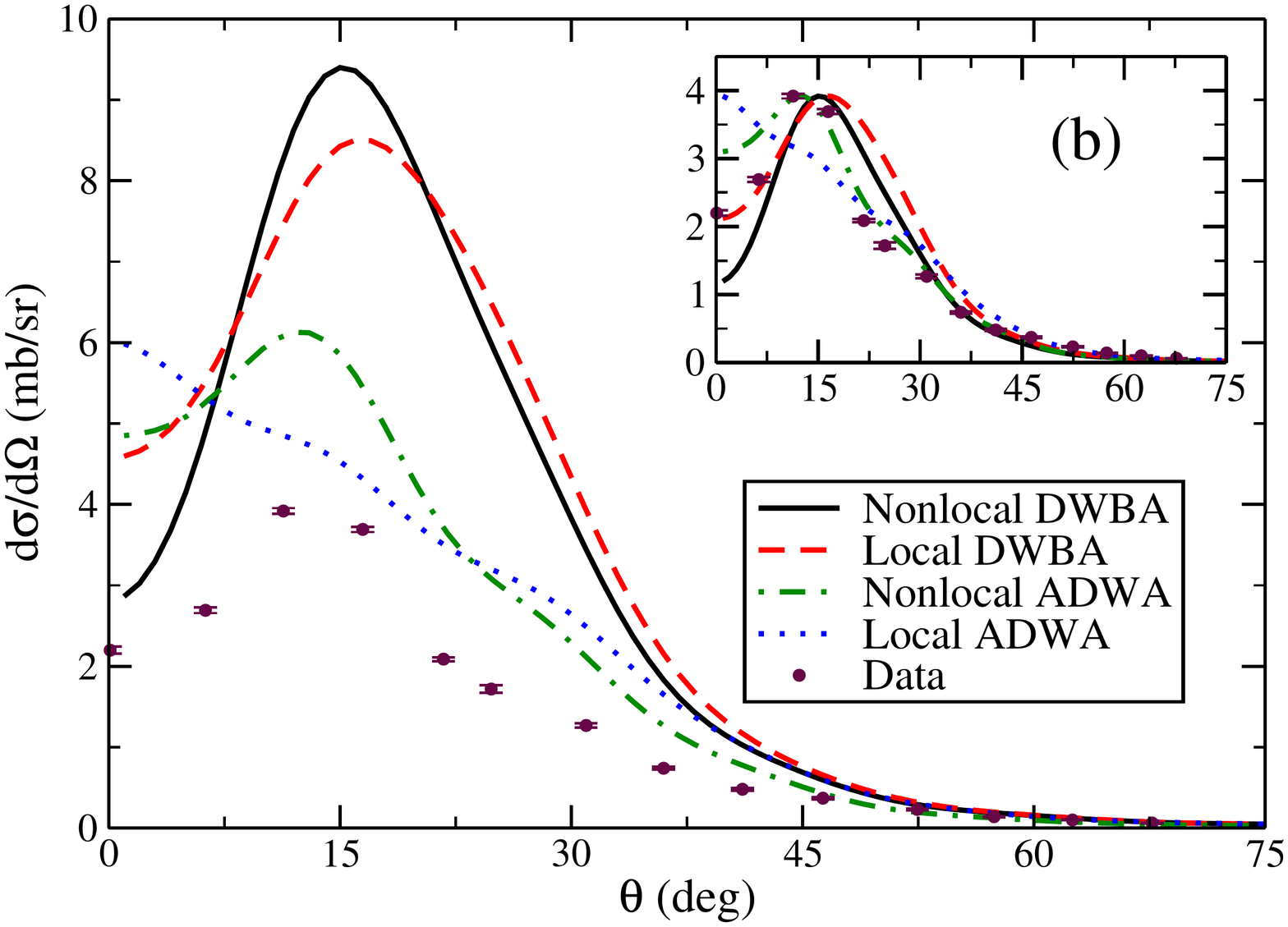}
\caption{(Color online) Angular distributions for $^{48}$Ca(d,n)$^{49}$Sc at 20 MeV (a) and 79 MeV with data from ~\cite{48Ca_data} (b). The inset shows the theoretical distributions normalized to the peak of the data.}
\label{transfer-ca}
\end{figure}

\begin{figure}[t]
\center
\includegraphics[scale=0.27]{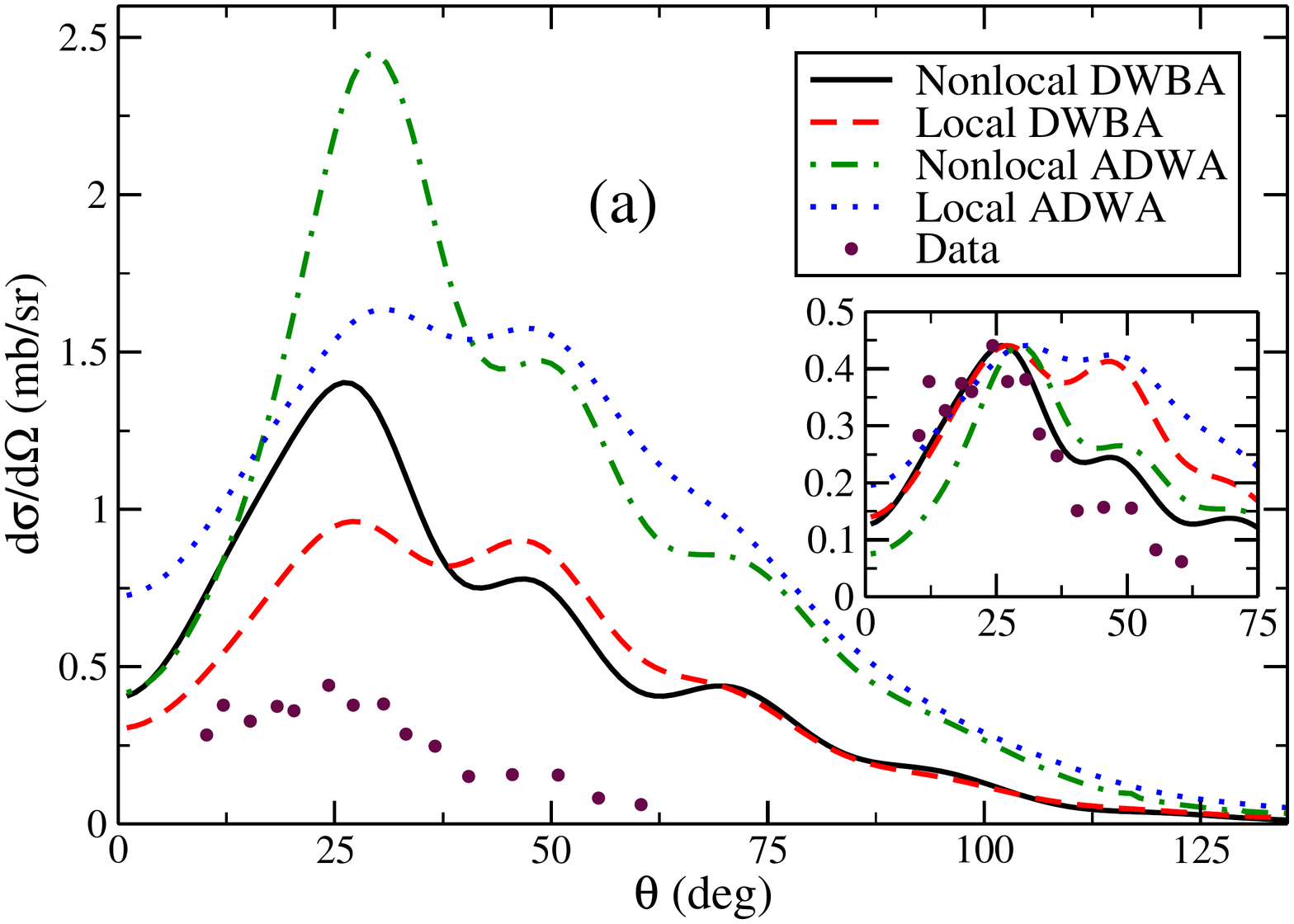}
\includegraphics[scale=0.27]{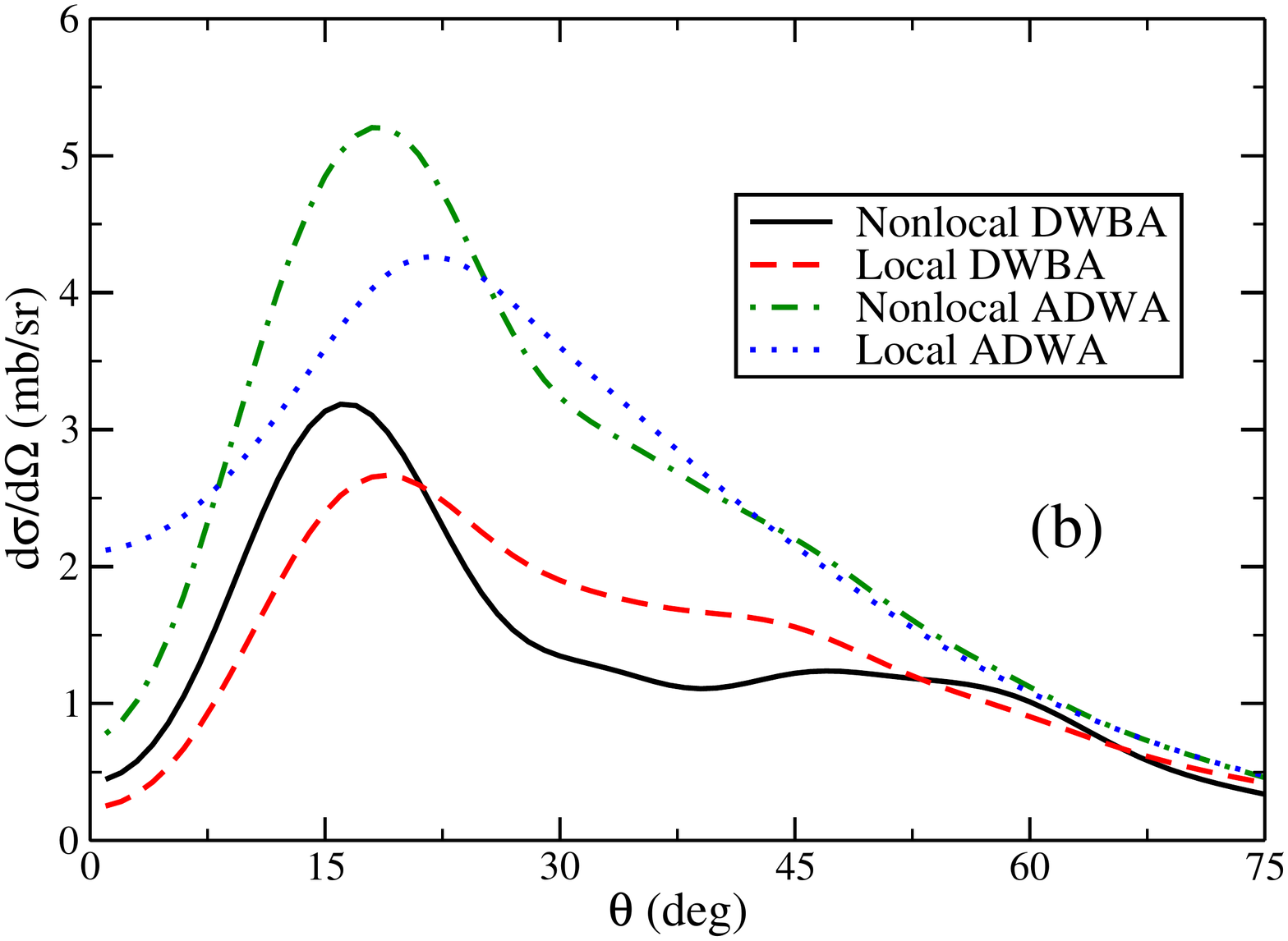}
\caption{(Color online) Angular distributions for $^{208}$Pb(d,n)$^{209}$Bi at 25 MeV with data from ~\cite{208Pb_data} (a) and 50 MeV (b). The inset shows the theoretical distributions normalized to the peak of the data.}
\label{transfer-pb}
\end{figure}

We first consider the effects of nonlocality in the exit channel, within DWBA, as done in \cite{Titus_prc2014}. 
Figs. \ref{transfer-ca} and \ref{transfer-pb} are an illustration of the effects obtained. In these figures we show the results when including nonlocality only in the exit channel (solid black line) and compare it with the corresponding local calculation (red dashed line). As in previous studies, nonlocality has a marked effect not only on the magnitude of the cross section but also on the shape. Overall, nonlocality in the final state increases the cross section, an effect that is more pronounced at lower energies. This effect is completely justified by the larger ANC $C_{nloc}$ as compared to $C_{loc}$ (see Table I).

We then also include nonlocality in the entrance channel, within ADWA, as done in \cite{Titus_prc2015}.  Figs. \ref{transfer-ca} and \ref{transfer-pb}  also contain the ADWA results: the green dot-dashed line includes nonlocality consistently in the entrance and exit channel and the blue dotted line is the corresponding ADWA local result.  Again, nonlocality affects both shape and magnitude of the differential cross sections. The effect of nonlocality in the deuteron channel is particularly strong for the heavier targets, consistent with the result from \cite{Titus_prc2015}. As observed in \cite{Titus_prc2015},  our (d,n) results using the first order DWBA can differ significantly from the distributions predicted within ADWA. This demonstrates the importance of including deuteron breakup explicitly in the reaction mechanism.

 In Figs. \ref{transfer-ca}(b) and \ref{transfer-pb}(a) we show the results of calculations at slightly different energies than those in our systematic study for the purpose of comparison to data from ~\cite{48Ca_data} and ~\cite{208Pb_data} respectively. The insets of these Figs. show the theoretical distributions normalized to the data at the peak,  to better compare the angular dependences.  Because the Perey and Buck potential, created in the sixties, was only fit to two data sets, it is not expected that it will do well reproducing the magnitude and details of the experimental angular distribution for a wide range of targets and energies. Nevertheless, as seen in Figs. \ref{transfer-ca}(b) and \ref{transfer-pb}(a), both DWBA and ADWA provide a reasonable qualitative  description of the process. Note that it is not the purpose of this work to analyse this data and extract structure information, but rather to unveil the effects on the transfer cross section of using nonlocal optical potentials, instead of the typical local potentials. So far, interpretation of data is done with local potentials, e.g. \cite{Anu_epja2016}.
 
In Fig.  \ref{transfer-pb}(a) DWBA provides a better descripion of the data than ADWA. This may be due to the assumption in ADWA that the excitation energy of the deuteron is small compared to the beam energy. For a detailed quantitative description of $^{208}$Pb(d,n)$^{209}$Bi at 25 MeV, one should use a model that includes deuteron breakup to all orders without making the adiabatic approximation, such a the Faddeev method. Currently, Faddeev calculations for heavy target are not feasible.  However, since in both DWBA and ADWA nonlocality effects are large, we expect that it will also hold true for more advanced reaction theories.

To quantify the nonlocal effects, we have summarized in Tables \ref{dwba20},\ref{dwba50},\ref{adwa20}, and \ref{adwa50} the percentage difference obtained at the peak of the cross section when nonlocality is included relative to the corresponding calculations when only local interactions are used:
\begin{eqnarray}\label{eq-tmatrix}
\Delta_{xs}=100 (\sigma_{nloc}(\theta_{peak}) - \sigma_{loc}(\theta_{peak}))/\sigma_{loc}(\theta_{peak} ) .
\label{diff-xs}
\end{eqnarray}

Focusing first on the effects seen within DWBA, it is clear that most of the effects we found are due to the effect of nonlocality in the bound state. This very strong sensitivity is largest for the lowest energy, as one would expect given that at 20 MeV the reaction is more peripheral. These large differences in the cross sections would mostly disappear if the ANC were constrained in our calculations. At higher energies, the magnitude of the nonlocal effect is the same as for the (d,p) reactions: up to $30$ \%.
\begin{table}[t]
\centering
\begin{tabular}{l r r r}
\hline
Reaction &  bound (\%) & scatt(\%) & total (\%) \\
\hline
$^{16}$O(d,n)$^{17}$F&  22.01		&-7.88	&      12.75\\
$^{40}$Ca(d,n)$^{41}$Sc& 40.30	&-1.66	&	39.46\\
$^{48}$Ca(d,n)$^{49}$Sc&  33.89	&-11.36	&	21.91\\
$^{126}$Sn(d,n)$^{127}$Sb&  50.83	&-7.08	&	52.43\\
$^{132}$Sn(d,n)$^{133}$Sb&  54.15	&-14.77	&	36.63\\
$^{208}$Pb(d,n)$^{209}$Bi&  64.64	&-13.52	&	53.21\\
\hline
\end{tabular}
\caption{Effects of nonlocality in (d,n) at 20 MeV within DWBA: percent differences of the cross section at  the peak of the angular distributions  including nonlocality relative to the cross section at the peak of the distribution when only local interactions are used: including nonlocality only in the bound state (bound), including nonlocaltity only in the scattering state (scatt) and including nonlocality in both states (total).}
\label{dwba20}
\end{table}
\begin{table}[t]
\centering
\begin{tabular}{l r r r}
\hline
Reaction &  bound (\%) & scatt (\%) & total (\%) \\
\hline
$^{16}$O(d,n)$^{17}$F	&   34.76	&-6.09&	27.39\\
$^{40}$Ca(d,n)$^{41}$Sc	&   19.33	&-9.43	&9.23\\
$^{48}$Ca(d,n)$^{49}$Sc	&   21.21	&-11.11	&9.53\\
$^{126}$Sn(d,n)$^{127}$Sb&   36.48&	-11.51	&24.29\\
$^{132}$Sn(d,n)$^{133}$Sb&   31.81	&-11.86	&20.72\\
$^{208}$Pb(d,n)$^{209}$Bi&   33.21	&-14.18	&19.51\\
\hline
\end{tabular}
\caption{Effects of nonlocality in (d,n) at 50 MeV within DWBA: percent differences of the cross section at  the peak of the angular distributions  including nonlocality relative to the cross section at the peak of the distribution when only local interactions are used: including nonlocality only in the bound state (bound), including nonlocaltity only in the scattering state (scatt) and including nonlocality in both states (total).}
\label{dwba50}
\end{table}

We next turn to the calculations within ADWA.  As seen in previous studies, the effects of nonlocality in the deuteron channel can be important, but effects appear to be less important in (d,n) than in (d,p). The effects in the entrance channel can cancel the effects in the exit channel. We find that, while at 20 MeV the overall effect can be very large (nearly a factor of 2), at the higher energy, the percentage difference only goes up to $\approx 20$ \%. The total percentage  differences in ADWA are in general smaller than those found in DWBA, particularly for the heavier targets, underlining the importance of consistently including nonlocality in the deuteron channel.

\begin{table}[t]
\centering
\begin{tabular}{l r r r}
\hline
Reaction &  entrance (\%) &  exit (\%) & total (\%) \\
\hline
$^{16}$O(d,n)$^{17}$F&  0.10&29.81&26.26\\
$^{40}$Ca(d,n)$^{41}$Sc& -3.19	&54.83&	43.42\\
$^{48}$Ca(d,n)$^{49}$Sc&  20.77	&26.73&	49.73\\
$^{126}$Sn(d,n)$^{127}$Sb&  20.29	&72.93&	95.01\\
$^{132}$Sn(d,n)$^{133}$Sb&  0.23	&29.63&	46.55\\
$^{208}$Pb(d,n)$^{209}$Bi& -13.43&	35.07&	25.18\\
\hline
\end{tabular}
\caption{Effects of nonlocality in (d,n) at 20 MeV within ADWA: percent differences of the cross section at  the peak of the angular distributions  including nonlocality relative to the cross section at the peak of the distribution when only local interactions are used:  including nonlocality only in the entrance channel (entrance), including nonlocality only in the exit channel (exit)  and including nonlocality in both (total).}
\label{adwa20}
\end{table}
\begin{table}[t]
\centering
\begin{tabular}{l r r r}
\hline
Reaction & entrance (\%) &  exit (\%) & total (\%) \\
\hline
$^{16}$O(d,n)$^{17}$F&   -1.10	&12.79&	14.16\\
$^{40}$Ca(d,n)$^{41}$Sc& 12.26	&17.82	&23.82\\
$^{48}$Ca(d,n)$^{49}$Sc&  -3.06	&-3.49	&-3.27\\
$^{126}$Sn(d,n)$^{127}$Sb&   -2.04&	8.26	&19.31\\
$^{132}$Sn(d,n)$^{133}$Sb&   -2.35	&4.01	&13.00\\
$^{208}$Pb(d,n)$^{209}$Bi&   9.33	&-0.63	&22.07\\
\hline
\end{tabular}
\caption{Effects of nonlocality in (d,n) at 50 MeV within ADWA: percent differences of the cross section at  the peak of the angular distributions  including nonlocality relative to the cross section at the peak of the distribution when only local interactions are used:  including nonlocality only in the entrance channel (entrance), including nonlocality only in the exit channel (exit)  and including nonlocality in both (total).}
\label{adwa50}
\end{table}

Because the final state in most reactions in our systematic study is not the same in (d,p) and (d,n), we performed additional test-studies, whereby both reactions populated the same final state and had the same separation energy. The conclusions from these additional calculations are clear: it is the Coulomb force in the final bound state that introduces the large differences we found in the magnitude of the nonlocal effects.
The effects of nonlocality in the nucleon and deuteron scattering states are in all similar in (d,p) and (d,n). However, the additional repulsive barrier caused by the Coulomb in bound-proton final state, produces a larger sensitivity to nonlocality. Constrains on the ANC of the final bound state would very much reduce the large dependences found. 

%%%%%%%%%%%%%%%%%%%%%%%%%%%%%%%%%%%%%%%%%%%%%%%%%%%%%%%%%%%%%%%%%%%%%%%%%%%%%%%%%%%%%%%%%%%%%%%%%%%%%%%%%%%%%%%%%

\section{Summary and Conclusions}
\label{conclusions}

In this work we explore the effects of nonlocality in (d,n) reactions. Our systematic study includes (d,n) reactions on $^{16}$O, $^{40}$Ca, $^{48}$Ca, $^{126}$Sn, $^{132}$Sn, and $^{208}$Pb at $20$ and $50$ MeV. We use both the distorted wave Born approximation and the adiabatic wave approximation to compare the results on (d,n) with those on (d,p) (\cite{Titus_prc2014} and \cite{Titus_prc2015} respectively). For a meaningful comparison with \cite{Titus_prc2014,Titus_prc2015} we use the nonlocal interaction, namely Perey and Buck \cite{Perey_np1962}, and impose the same physical constrains in generating the local interactions. For the scattering, NA interaction, a local phase equivalent potential was obtained by fitting the elastic scattering generated from the corresponding nonlocal potential.  The local and nonlocal bound states reproduced the same experimental binding energies. Effects of nonlocality are determined by comparing the transfer cross sections using the nonlocal optical potentials and the phase equivalent local interactions.

Just as in the (d,p) case, DWBA calculations show that nonlocality in the final bound state increased the cross sections for (d,n), while nonlocality in the final scattering state, reduces those cross sections. However the effect of the final bound state is much larger than that of the scattering state, and therefore overall cross sections are increased due to nonlocality. This increase is substantially larger in (d,n) than in (d,p) because the presence of the Coulomb interaction in the bound state increases its sensitivity to nonlocality. This is the most marked difference between (d,n) and (d,p).

For the ADWA results, nonlocality in the deuteron channel is insignificant for light targets but can have a substantial effect on the cross section  for intermediate mass and heavy targets (just as we observed in the (d,p) case). It is also very dependent on the particular characteristics of the final state being populated. The interplay of the effects of nonlocality in the entrance and the exit channel produce a total percentage difference at the peak of the angular distributions that can go up to a factor of two for the lower energy, while it is around $20$\% for the higher energy case.

Given the very strong effect of nonlocality on the asymptotic normalization coefficient for proton bound states, it is important that future analyses of (d,n) be performed under a physical constrain on the ANC, which may come from another peripheral reaction measurement.

%%%%%%%%%%%%%%%%%%%%%%%%%%%%%%%%%%%%%%%%%%%%%%%%%%%%%%%%%%%%%%%%%%%%%%%%%%%%%%%%%%%%%%%%%%%%%%%%%

\begin{center}
\textbf{ACKNOWLEDGMENTS}
\end{center}

We are grateful to Charlotte Elster for  useful discussions.
This work was supported by the National Science
Foundation under Grants No. PHY-1068571 and PHY-1403906 and the
Department of Energy under Contract No. DE-FG52-
08NA28552.

%%%%%%%%%%%%%%%%%%%%%%%%%%%%%%%%%%%%%%%%%%%%%%%%%%%%%%%%%%%%%%%%%%%%%%%%%%%%%%%%%%%%%%%%%%%%%%%%%

\bibliography{nonlocal_transfer}

\end{document}